\documentclass[10pt,conference]{IEEEtran}
\usepackage{cite}
\usepackage{amsmath,amssymb,amsfonts}
\usepackage[linesnumbered,ruled,vlined]{algorithm2e}
\usepackage{graphicx}
\usepackage{textcomp}
\usepackage{tikz}
\usepackage{xcolor}
\usepackage[hyphens]{url}
\usepackage{fancyhdr}
\usepackage{hyperref}
\usepackage{marvosym}

\newcommand*\circled[1]{\tikz[baseline=(char.base)]{
            \node[shape=circle,fill, inner sep=0pt, minimum width=0.3cm] (char) {\textcolor{white}{#1}};}}

\newcommand{\hpcayear}{2025}

\newcommand{\hpcasubmissionnumber}{306}
\title{Make LLM Inference Affordable to Everyone: Augmenting GPU Memory with NDP-DIMM}

\def\hpcacameraready{} 

\newcommand\hpcaauthors{Lian Liu\textsuperscript{$1, 2, 3, \dagger$}, Shixin Zhao\textsuperscript{$1, 2, \dagger$}, Bing Li\textsuperscript{4}, Haimeng Ren\textsuperscript{1,5}, Zhaohui Xu\textsuperscript{1,5}, \\ Mengdi Wang\textsuperscript{1,2}, Xiaowei Li\textsuperscript{1,2,3}, Yinhe Han\textsuperscript{1,2}, and Ying Wang\textsuperscript{1,2, \Letter}}
\newcommand\hpcaaffiliation{Institute of Computing Technology, Chinese Academic of Sciences\textsuperscript{$1$}, \\ University of Chinese Academy of Sciences \textsuperscript{$2$}, Zhongguancun Laboratory\textsuperscript{$3$}, \\
Institute of Microelectronics, Chinese Academy of Sciences\textsuperscript{$4$}, \\
School of Information Science and Technology, ShanghaiTech University\textsuperscript{$5$}}
\newcommand\hpcaemail{ \{liulian211, zhaoshixin18\}@mails.ucas.ac.cn \quad libing2024@ime.ac.cn \quad \{renhm2022, xuzhh12022\}@shanghaitech.edu.cn \\ \{wangmengdi, lxw, yinhes, \textcolor{blue}{wangying2009}\}@ict.ac.cn}

\def\sec{Section~}

 \def\fig{Figure~}

\def\tab{Table~}

\def\eq{Equation~}
\definecolor{mygray}{gray}{0.9}

\usepackage{dsfont}

\usepackage{xspace}

\def\@onedot{\ifx\@let@token.\else.\null\fi\xspace}

\def\name{Hermes}

\newcommand\update[1]{\textcolor{black}{#1}}

\usepackage{multirow}
\usepackage{colortbl}
\usepackage{tabularx}
\usepackage[ruled]{algorithm2e}

\author{
  \ifdefined\hpcacameraready
    \IEEEauthorblockN{\hpcaauthors{}}
      \IEEEauthorblockA{
        \hpcaaffiliation{} \\
        \hpcaemail{}
      }
  \else
    \IEEEauthorblockN{\normalsize{HPCA \hpcayear{} Submission
      \textbf{\#\hpcasubmissionnumber{}}} \\
      \IEEEauthorblockA{
        Confidential Draft \\
        Do NOT Distribute!!
      }
    }
  \fi 
}

\fancypagestyle{camerareadyfirstpage}{%
  \fancyhead{}
  
  \fancyhead[C]{
    \ifdefined\aeopen
    \parbox[][12mm][t]{13.5cm}{\hpcayear{} IEEE International Symposium on High-Performance Computer Architecture (HPCA)}    
    \else
      \ifdefined\aereviewed
      \parbox[][12mm][t]{13.5cm}{\hpcayear{} IEEE International Symposium on High-Performance Computer Architecture (HPCA)}
      \else
      \ifdefined\aereproduced
      \parbox[][12mm][t]{13.5cm}{\hpcayear{} IEEE International Symposium on High-Performance Computer Architecture (HPCA)}
      \else
      \parbox[][0mm][t]{13.5cm}{\hpcayear{} IEEE International Symposium on High-Performance Computer Architecture (HPCA)}
    \fi 
    \fi 
    \fi 
    \ifdefined\aeopen 
      \includegraphics[width=12mm,height=12mm]{ae-badges/open-research-objects.pdf}
    \fi 
    \ifdefined\aereviewed
      \includegraphics[width=12mm,height=12mm]{ae-badges/research-objects-reviewed.pdf}
    \fi 
    \ifdefined\aereproduced
      \includegraphics[width=12mm,height=12mm]{ae-badges/results-reproduced.pdf}
    \fi
  }
  \fancyfoot[C]{}
}
\begin{document}
\maketitle
    
\ifdefined\hpcacameraready 
  \thispagestyle{camerareadyfirstpage}
  \pagestyle{empty}
\else
  \thispagestyle{plain}
  \pagestyle{plain}
\fi

\newcommand{\hpcaheight}{0mm}
\ifdefined\eaopen
\renewcommand{\hpcaheight}{12mm}
\fi

\begin{abstract}
The billion-scale Large Language Models (LLMs) necessitate deployment on expensive server-grade GPUs with large-storage HBMs and abundant computation capability. As LLM-assisted services become popular, achieving cost-effective LLM inference on budget-friendly hardware becomes the current trend. This has sparked extensive research into relocating LLM parameters from expensive GPUs to external host memory. However, the restricted bandwidth between the host and GPU memory limits the inference performance of existing solutions.

This work introduces Hermes, a budget-friendly system that leverages the near-data processing units (NDP) within commodity DRAM DIMMs to enhance the performance of a single consumer-grade GPU, achieving efficient LLM inference. We recognize that the inherent activation sparsity in LLMs naturally divides weight parameters into two categories, termed ``hot" and ``cold" neurons, respectively. Hot neurons, which consist of only approximately 20\% of all weight parameters, account for 80\% of the total computational load. In contrast, cold neurons make up the other 80\% of parameters but are responsible for just 20\% of the computational workload. Leveraging this observation, we propose a heterogeneous computing strategy: mapping hot neurons to a single computation-efficient GPU without large-capacity HBMs, while offloading cold neurons to NDP-DIMMs, which offer large memory size but limited computation capabilities. In addition, the dynamic nature of activation sparsity necessitates a real-time partition of hot and cold neurons and adaptive remapping of cold neurons across multiple NDP-DIMM modules. To tackle these issues, we introduce a lightweight predictor that ensures optimal real-time neuron partition and adjustment between GPU and NDP-DIMMs. Furthermore, we utilize a window-based online scheduling mechanism to maintain load balance among multiple NDP-DIMM modules. In summary, Hermes facilitates the deployment of LLaMA2-70B on consumer-grade hardware at a rate of 13.75 tokens/s and realizes an average 75.24$\times$ speedup over the state-of-the-art offloading-based inference system on popular LLMs.

\end{abstract}

\maketitle 
\thispagestyle{empty}
\pagestyle{empty}

\def\thefootnote{$\dagger$}\footnotetext{Both authors contributed equally to this research}\def\thefootnote{\arabic{footnote}}
\def\thefootnote{\Letter}\footnotetext{Corresponding author}\def\thefootnote{\arabic{footnote}}

\section{Introduction}

Large Language Models (LLMs) have gained significant importance and widespread attention. Open-source models like OPT, LLaMA, and Qwen series~\cite{zhang2022opt, qwen2, touvron2023llama2}, as well as proprietary models such as GPT-4 and Claude~\cite{achiam2023gpt, claude3}, exhibit remarkable performance in a variety of tasks including code generation~\cite{chen2021evaluating, Copilot}, machine translation~\cite{le2023bloom, jiang2023mistral}, and chatbots~\cite{chatgpt2023, bard2023}, etc. Nevertheless, extremely powerful LLMs with billions of parameters often require server-grade GPUs with large-capacity HBMs, making them cost-prohibitive for many applications. For example, deploying LLaMA2-70B locally using TensorRT-LLM~\cite{tensorrt-llm} requires five NVIDIA A100-40GB-SXM4 GPUs, totaling over \$50,000.

\begin{figure}
    \centering
    \includegraphics[width=\linewidth]{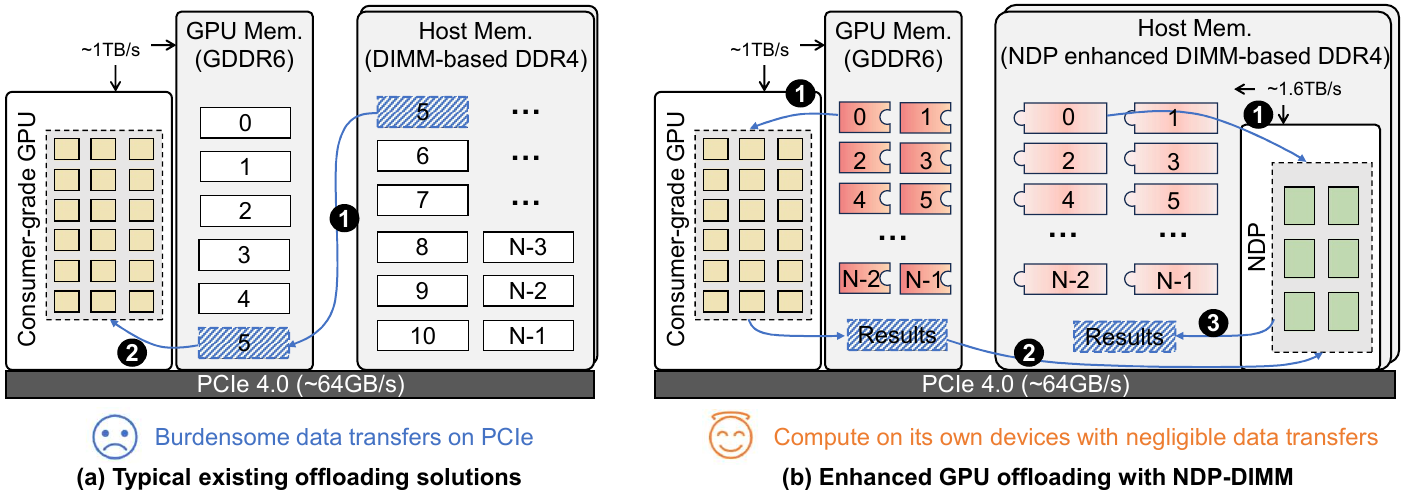}
    \vspace{-0.3cm}
    \caption{(a) Existing offloading solutions view host memory as the augmented memory, but cause burdensome data transfer on PCIe. (b) Partitioning the weight matrix in each layer, and utilizing NDP-DIMMs to handle poor computation intensity parts, only introduces negligible data transfer.}
    \label{fig:offloading}
\vspace{-0.3cm}
\end{figure}

To investigate the development of cost-effective LLM inference systems, researchers have shifted their focus to more budget-friendly hardware, such as consumer-grade GPUs. Despite these GPUs' significant computation capability, such as 1321 Tensor TOPS in NVIDIA RTX 4090, they suffer from limited graphic memory size. This limitation renders them unsuitable for deploying LLMs with billions of parameters. To this end, researchers use offloading strategies~\cite{jain2022hugging, rasley2020deepspeed, sheng2023flexgen}, transferring large portions of LLM parameters to DIMM (Dual-Inline Memory Module)-based host memory. As depicted in \fig \ref{fig:offloading}a, existing offloading solutions view host memory as the augmented memory space for GPUs to enable LLMs, and parameters stored in host memory need to be accessed via PCIe. This results in substantial data transfers on PCIe. However, due to more than $15 \times$ bandwidth gap between the PCIe and the internal GPU memory, about 99\% of the overall LLM runtime in these offloading solutions is attributed to the data transfers on PCIe. 

It is essential to minimize the data loading for weight parameters to ease the burden on PCIe. Thus, existing works~\cite{liu2023deja, song2023powerinfer, xue2024powerinfer} utilize the activation sparsity to reduce the required data loading.
Since the activation functions such as ReLU in LLMs can zero out specific activation values, the corresponding parameters that are expected to be computed with these zero activations do not need to be loaded either, as illustrated in \fig \ref{fig:pruning-network}.  According to the activation sparsity, weight parameters in LLMs can be further categorized into hot and cold neurons\footnote{This paper defines a neuron as a specific row/column in a weight matrix, and neurons will not be activated when associated with zero activations.}. Our evaluation indicates that around 20\% of neurons, referred to as ``hot neurons'', are responsible for 80\% of the computations, whereas the remaining 80\% of neurons, known as ``cold neurons'', account for only 20\% of the computations. This suggests that the computation intensity of hot neurons is $16 \times$ higher than that of cold neurons. Consequently, it is natural to store hot neurons in GPU memory and offload cold neurons to host memory to effectively mitigate data loading costs~\cite{liu2023deja}. Despite these optimizations, data transfers on PCIe still dominate the inference procedure, accounting for 90\% of the total inference latency of OPT-66B, as they constitute a large part of the total LLM work-set.

According to our observation, the cold neurons offloaded on host memory require large storage but have poor computation intensity. As a result, we are motivated to utilize near-data processing (NDP) units based on DRAM DIMMs to provide the least-required computation capability for cold neurons to avoid their movement. As illustrated in \fig \ref{fig:offloading}b, we can leverage the NDP units and GPU cores to conduct computations for cold and hot neurons, respectively. As the computation results only take a few KBs, the data transfer cost in step \circled{2} is negligible. Note that we use NDP-DIMMs, instead of high-performance but expensive alternatives such as HBM-PIM and AiM~\cite{park2024attacc, heo2024neupims, gao2017tetris, cong2017aim}, as the augmented memory to build the budget-friendly system for local deployment.

Yet, attaining high-performance but affordable LLM inference using a basic NDP-DIMM enhanced GPU system is challenging due to the limited computational resources in NDP-DIMMs. Two primary challenges must be resolved:


\textbf{1. Deciding the optimal neuron partition. }
First, the criteria for dividing hot and cold neurons between GPU and NDP-DIMMs are crucial for computational efficiency. For instance, if only the least active neurons are predicted as ``hot", this will stress the limited GPU memory size. Conversely, allocating frequently activated neurons to the ``cold" region will burden the computation-limited NDP-DIMMs with excessive computation. Therefore, determining the optimal neuron partition strategy is essential. However, due to the input-specific nature, the hot/cold neuron partition cannot be completely predetermined. It necessitates an accurate but lightweight online prediction to achieve real-time adjustment for hot/cold neuron partition with minimal migration cost. 


\textbf{2. Exploiting the limited computation capability of multiple NDP-DIMMs. }
In contrast to the provided hundreds of TFLOPS of a single GPU, the computation capability is constrained to hundreds of GFLOPS~\cite{asghari2016chameleon,zhou2023dimm,devaux2019true,kim2021aquabolt} on NDP-DIMMs. Consequently, even are used to process the infrequently activated neurons, NDP-DIMMs still bottleneck the inference performance. Thus, it is crucial to fully unleash NDP units for efficient computing. Specifically, as we need to use multiple DIMMs together to support the large-scale LLMs, computational loads on each NDP-DIMM are expected to be balanced. However, due to the dynamics of activated neurons, some NDP-DIMMs are overburdened while others remain underutilized during inference. Therefore, the main challenge is to achieve online scheduling for computational load balance among NDP-DIMMs.


To address the aforementioned challenges, we introduce \name, an innovative and budget-friendly inference system that uses NDP-DIMMs to enhance both the memory capacity and processing capability of a single consumer-grade GPU. On one hand, we address the optimal neuron partition in two phases. First, we formalize the problem as an integer linear programming (ILP) issue and employ an offline solver to help determine the optimal partition based on the profiled data. Then, utilizing the distinct distribution patterns of hot and cold neurons, we develop a lightweight online predictor to manage online cold/hot neuron partition. This approach bypasses the expensive MLP-based predictor used in prior studies~\cite{song2024prosparse, song2024turbo, xue2024powerinfer}, enabling real-time migration of hot and cold neurons. On the other hand, to address load imbalance issues among multiple NDP-DIMMs, we exploit the token-wise similarity inherent in LLM. In detail, we propose a window-based online scheduling strategy, which utilizes the neuron activity of adjacent tokens to online remap cold neurons across multiple NDP-DIMMs, achieving load balance.


In a nutshell, our contributions are as follows:
\begin{enumerate}
    \item We propose a novel system, \name, which takes advantage of the cold/hot distribution in LLM inference and augments consumer-grade GPU with NDP-DIMMs to achieve high-performance and economic LLM inference.
    \item We propose a two-step solution to achieve the optimal cold/hot neuron partition for \name. We first formulate an ILP problem and utilize an offline solver to find the original optimal partition, and further implement a lightweight online predictor to guide the online migration of hot and cold neurons during LLM inference. 
    \item We develop a window-based online scheduling strategy to achieve load balance among multiple computation-limited NDP-DIMMs, effectively improving the overall hardware utilization.
    \item Compared to existing offloading-based inference systems FlexGen and Deja Vu, \name~achieves a speedup of $148.98 \times$ and $75.24 \times$, respectively. 
\end{enumerate}

\section{Background}

\subsection{LLM Inference \& Architecture}\label{sec:llm-procedure}

\begin{figure}
    \centering
    \includegraphics[width=0.98\linewidth]{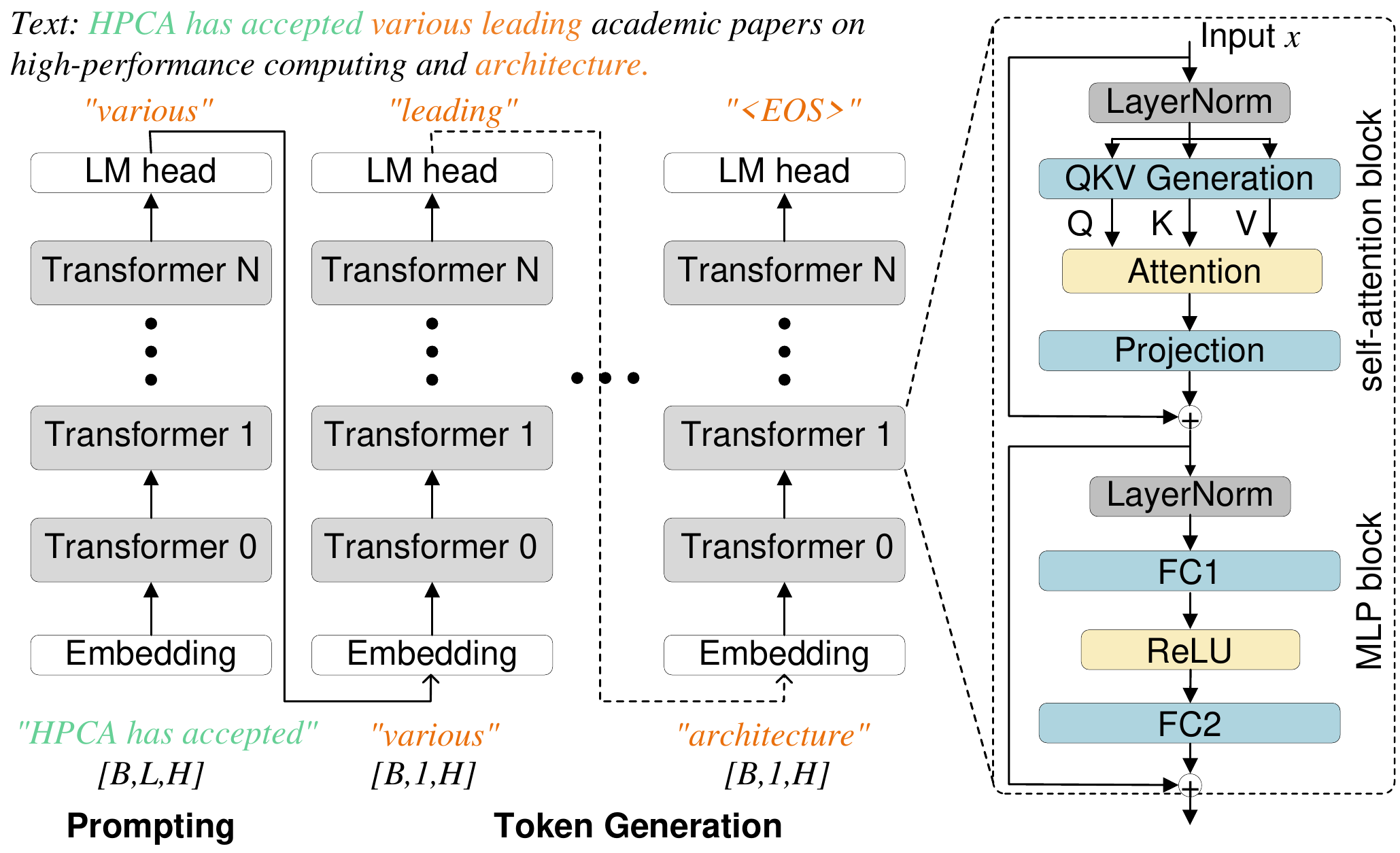}
    \vspace{-0.3cm}
    \caption{The LLM inference procedure and architecture.}
    \label{fig:llm-execution}
\vspace{-0.3cm}
\end{figure}

As shown in \fig \ref{fig:llm-execution}, the inference procedure of transformer-based LLMs comprises two stages: prompting and token generation. During the prompting stage, an input sequence is used to produce keys and values (KV cache) for each transformer layer in the LLM, and this is done just once per inference. In the token generation stage, previously generated tokens are used to update the KV cache and generate new tokens incrementally. This stage is executed multiple times, depending on the length of the output sequence. \update{Since token generation accounts for more than 90\% of the total runtime~\cite{li2024specpim}, this paper primarily focuses on optimizing inference efficiency in token generation.}


An LLM has multiple transformer layers, each containing a self-attention and an MLP block. In the self-attention block, input \(x\) is projected linearly to produce Q, K and V, processed by the attention operator to yield the attention result, and then computed by the projection layer for the MLP input. The MLP block includes fully connected (FC) layers and non-linear functions. For example, the OPT model uses two FC layers which are connected by one ReLU activation function.

\subsection{Activation Sparsity in LLMs}\label{sec:activation-sparsity}

\begin{figure}
    \centering
    \includegraphics[width=0.98\linewidth]{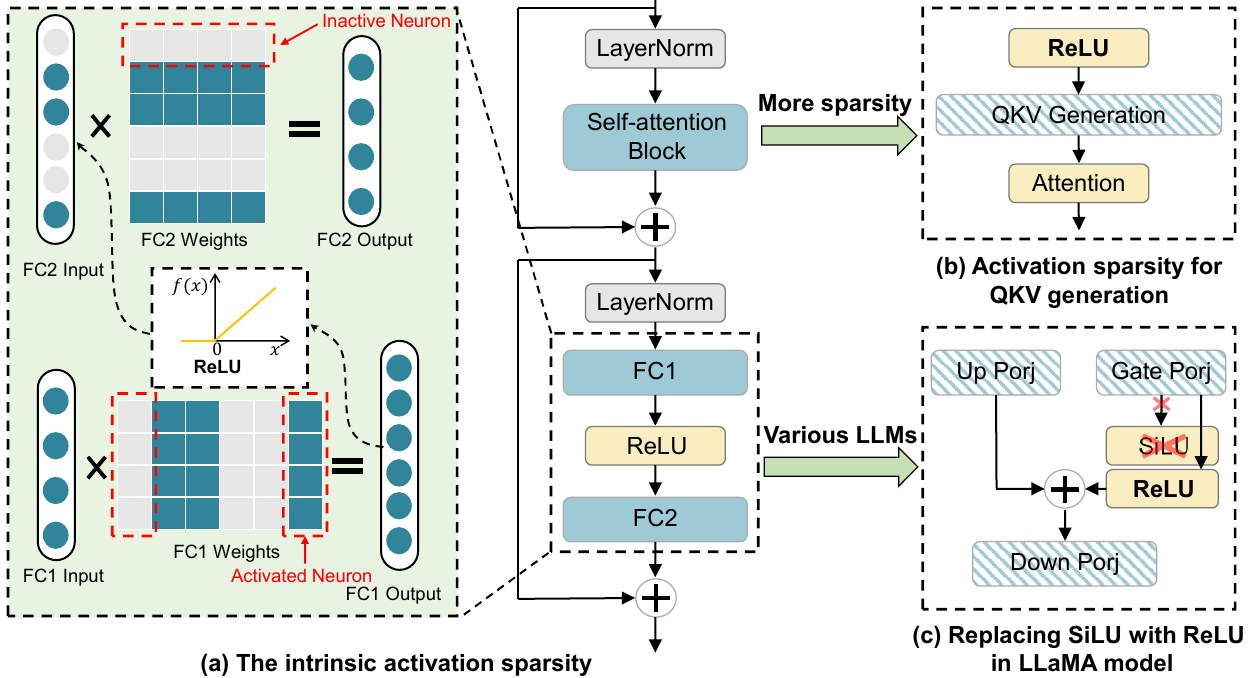}
    \vspace{-0.3cm}
    \caption{The inherent activation sparsity within certain LLMs is further enhanced to achieve higher sparsity across various LLMs.}
    \label{fig:pruning-network}
\vspace{-0.3cm}
\end{figure}

The activation function such as ReLU in the MLP block introduces the intrinsic activation sparsity to LLMs~\cite{liu2023deja, mirzadeh2023relu, song2024prosparse}. As shown in \fig \ref{fig:pruning-network}a, the ReLU function in the MLP block, can turn many activation values to zero, eliminating the need to load and compute these inactive neurons. As the red dashed box shows, a neuron in this paper represents a specific row or column within a weight matrix. For example, due to the ReLU function zeros out the 1st, 4th and 5th input values of the FC2 layer, the corresponding columns and rows in FC1 and FC2 weight matrix will not be activated. 

To further achieve activation sparsity on self-attention blocks, programmers insert ReLU functions before QKV generation~\cite{mirzadeh2023relu}, as illustrated in \fig \ref{fig:pruning-network}b. For LLMs that do not use ReLU as their activation function, such as LLaMA (SiLU) and Falcon (GELU)~\cite{touvron2023llama2, almazrouei2023falcon}, recent work has demonstrated that they can also be replaced by ReLU functions~\cite{mirzadeh2023relu, song2024prosparse}, as demonstrated in \fig \ref{fig:pruning-network}c. Previous studies~\cite{song2024turbo, zheng2024learn, song2024prosparse} also demonstrated that the activation sparsity within LLMs provides significant sparsity (ranging from 70\% to 90\%) with negligible accuracy degradation (less than 1\%).

\subsection{Offloading-based LLM Inference Systems}\label{sec:background-offloading}
Most existing LLM inference systems~\cite{tensorrt-llm, rasley2020deepspeed, yu2022orca, kwon2023efficient} require the use of expensive server-grade GPUs, which provide high-capacity HBM to store the large-scale LLM parameters. This limits their deployment to easily accessible and affordable hardware. Offloading is a viable technique to enable LLM inference on such commodity hardware~\cite{jain2022hugging}. For instance, a single consumer-grade GPU can leverage the host memory resources to perform inference of LLaMA2-70B~\cite{rasley2020deepspeed, huggingface-accelerate}. 

Existing offloading-based inference systems utilize host memory to extend the storage capacity of the GPU to accommodate LLMs. As long as there is sufficient host memory, this strategy can be used to perform inference on LLMs of various sizes. 
HuggingFace Accelerate~\cite{jain2022hugging} integrates offloading techniques from training systems by automatically mapping and partitioning weights into GPU and host memory respectively, only transferring the necessary parameters during inference. However, the characteristics of LLM inference are quite different from training~\cite{cho2024llmservingsim}, making it inefficient. To address this issue, FlexGen~\cite{sheng2023flexgen} provides a novel zig-zag offloading strategy to maximize the inference throughput within a limited PCIe bandwidth. This zig-zag scheduling strategy integrates multiple tokens into a block and overlaps the weight-loading cost during token processing within one block. For instance, it computes all the tokens in one block (e.g., more than 100 tokens) with the weights in layer $i$, while prefetching the weights in layer $i+1$ simultaneously. The burdensome block computation in one layer effectively overlaps the weight prefetching cost for the next layer, especially for the prefill phase which occupies multiple tokens even with a single batch. However, this method is unsuitable for local deployment scenarios, which only occupy limited batch sizes~\cite{hong2024flashdecoding++} during token generation.
Deja Vu~\cite{liu2023deja} further exploits \textit{activation sparsity} to perform LLM inference by predicting and loading only the activated neurons, thereby reducing data access and computation overhead. 
However, since the activated neurons are dynamic and cannot be pre-loaded into the limited consumer-grade GPU memory, data still need to be loaded from host memory, resulting in inference efficiency being bounded by PCIe bandwidth. 
Overall, while existing offloading solutions can effectively extend the storage capacity of inference systems to support larger LLMs, the low bandwidth data transfer of PCIe results in poor inference performance. 
 

\section{Motivations \& Challenges}

\subsection{Why NDP-DIMM Enhanced GPU?}


Offloading is essential for LLM inference on low-budget systems with a single consumer-grade GPU. However, as noted in Section \ref{sec:background-offloading}, even utilizing activation sparsity to reduce weight parameter access, the PCIe bandwidth remains the bottleneck. Thus, costly data transfers between extended memory and GPU must be minimized. However, simply offloading the corresponding computation of cold neurons on the host CPU~\cite{llama.cpp, song2023powerinfer} can only achieve a limited performance improvement, \update{as the host CPU can only access DRAM with limited improved bandwidth than PCIe (e.g., 89.6 GB/s vs. 64 GB/s)}. To this end, we choose to employ multiple NDP-DIMMs as the extended memory, as they offer comparable bandwidth and larger storage capacity than a single consumer-grade GPU. Need to mention that as a budget-friendly host memory solution, we do not consider high-performance but expensive HBM-PIM and AiM~\cite{cong2017aim, park2024attacc} in this study. Given the limited computation capability, only utilizing the processing units in NDP-DIMMs cannot boost the inference efficiency~\cite{wu2024pim}. Consequently, we are motivated to use NDP-DIMMs to enhance GPU for efficient LLM inference.

Our observation indicates that the activation sparsity within LLMs effectively partitions weight parameters into two distinct regions, which are ideally suited to consumer-grade GPU and NDP-DIMMs, respectively. Specifically, activation sparsity in LLMs follows a power-law distribution~\cite{xue2024powerinfer, song2023powerinfer}. About 20\% of neurons (\textit{hot neurons}) account for 80\% of computations, while 80\% (\textit{cold neurons}) handle only 20\%. Hot neurons, with $16 \times$ higher computation intensity, fit GPU memory, while cold neurons suit NDP-DIMMs. During inference, GPU can provide high computation capability for hot neurons and NDP-DIMMs enable the cold neurons computation in memory.

\subsection{Necessity of Hot/Cold Neuron Partition} \label{sec:similar}

Hot/cold neuron partition impacts the computational load on GPU/NDP-DIMMs, affecting the inference performance of the heterogeneous system. Due to the input-specific nature of activation sparsity, solely relying on the offline partition is insufficient. Our evaluation on LLaMA2-70B reveals significant dynamics in when the neuron will be activated (hereafter, neuron activity patterns) during inference. Approximately 52\% of the initialized hot neurons exhibit varied activity during inference. This variability in neuron behavior results in suboptimal performance with a fixed hot/cold partition, causing a $1.63\times$ degradation compared to an oracle (the theoretically optimal partition) scheme. Thus, we must dynamically predict and adjust the hot/cold neuron partition.

However, typical MLP-based predictors~\cite{liu2023deja, song2024prosparse, zhang2024relu, mirzadeh2023relu} for activation sparsity in LLMs are costly. For example, predicting the activated neurons in LLaMA-7B needs per-layer MLP-based predictors, requiring an extra 2GB storage and inducing 10\%-25\% inference runtime. Fortunately, the inherent locality of activation sparsity leads us to design a lightweight and accurate predictor for efficient online partition adjustments. To be specific, we found that activation sparsity in LLM inference shows considerable token-wise similarity and layer-wise correlation, worth exploiting.

\begin{figure}[t]
    \centering
    \includegraphics[width=0.98\linewidth]{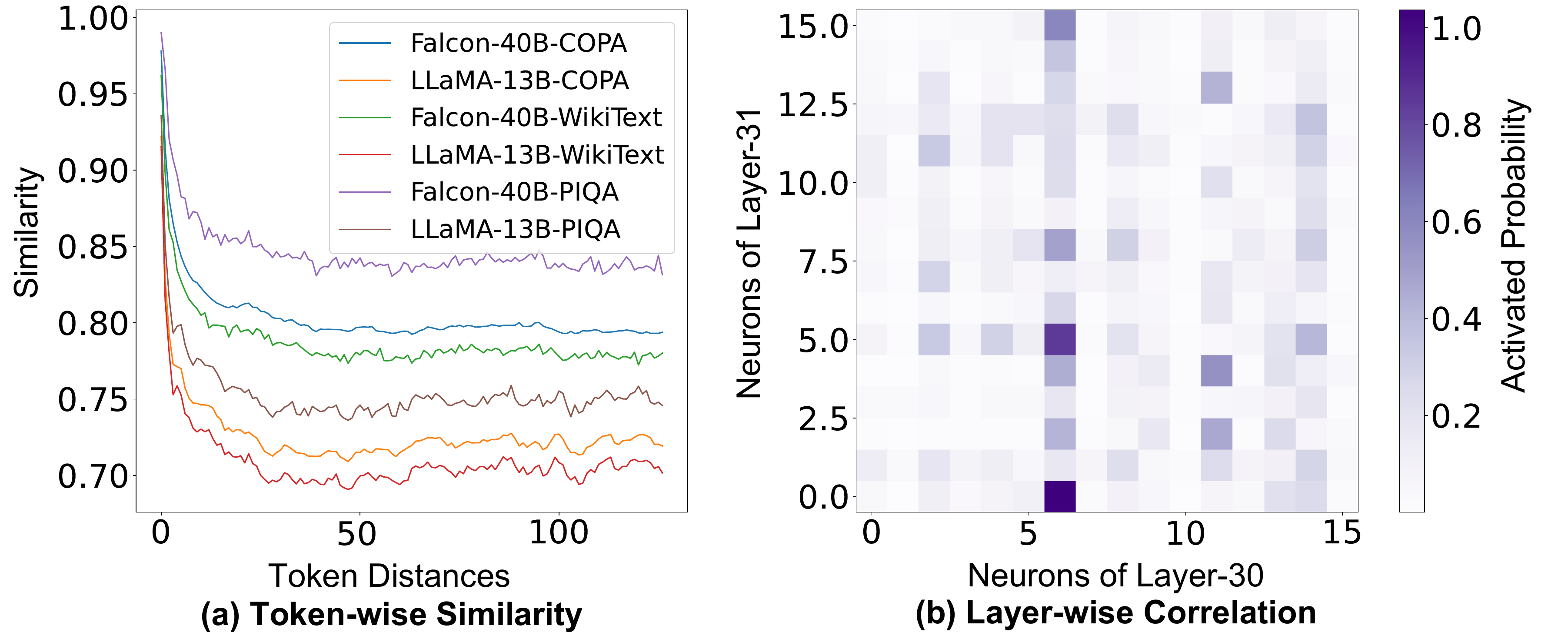}
    \vspace{-0.3cm}
   \caption{Distribution patterns for activation sparsity. (a) \update{The adjacent tokens enjoy high similarity on activated neurons for various models and datasets.} (b) The activated neurons between consecutive layers are highly correlated.}
    \label{fig:similarity}
\vspace{-0.3cm}
\end{figure}

\subsubsection{Token-wise Similarity}
We analyzed the similarity between tokens to explore the distribution characteristics of activation sparsity. \update{As shown in \fig \ref{fig:similarity}a, we evaluate the token-wise similarity for LLaMA-13B and Falcon-40B with multiple widely adopted datasets, including COPA~\cite{roemmele2011choice}, Wikitext2~\cite{merity2016pointer} and PIQA~\cite{bisk2020piqa}. As one can notice, the adjacent tokens have a higher distribution similarity than distant tokens. Specifically, the similarity between adjacent tokens exceeds 90\% (95\% for Falcon-40B), but drops to 70\% once the tokens' distance exceeds 10.} This indicates that in context, adjacent tokens often express similar meanings, leading to high similarity in their activity distribution. Additionally, we observe that when the distance between tokens exceeds 25, the distribution similarity almost no longer decreases, indicating that beyond a certain window size, the semantic correlation becomes weak and has less impact on the overall distribution.

\subsubsection{Layer-wise Correlation} 
We further observed that the distribution of activated neurons in two consecutive layers is highly correlated. As shown in \fig \ref{fig:similarity}b, when the 6th neuron in layer-30 of LLaMA-13B is activated, the probability of neurons 0 and 5 being activated in layer-31 exceeds 90\%. This suggests that we can use the results of the preceding layer to predict the distribution of activated neurons in the current layer.

Overall, the token-wise similarity and layer-wise correlation motivate us to design a lightweight online predictor based on historical activation information. According to the prediction results, we can online adjust the hot/cold neurons partition to effectively exploit the processing advantages of the consumer-grade GPU and NDP-DIMMs, respectively.


\subsection{Load Imbalance across Multiple NDP-DIMMs}
Due to the storage limitation of a single DIMM, multiple DIMMs are required to store all the neurons (weight parameters) in LLM. Specifically, one DIMM only stores portions of the neurons and the corresponding processing unit can only directly assess neurons in the DIMM with high internal bandwidth. However, due to the input-specific nature of activation sparsity, the computational load on each NDP-DIMM can be diverse. For example, when fixing the cold neuron distribution on multiple DIMMs for LLaMA-13B, the most overloaded NDP-DIMM will have 1.2-2.5$\times$ more computational load than others.

Therefore, we need an online scheduling strategy to remap the cold neuron across DIMMs to achieve load balance. Meanwhile, an efficient data transmission pathway among DIMMs is essential to help adjust the neuron placement. By optimizing neuron computation scheduling, we can minimize data transfers across NDP-DIMMs while ensuring balanced computational loads across DIMMs. This ensures that all parts of the system can maximize their performance.

In summary, the NDP-DIMM enhanced GPU approach effectively addresses the substantial data transfer overhead in offloading processes, providing a promising solution to improve LLM inference efficiency by leveraging the activation sparsity patterns inherent in LLMs.

\begin{figure*}
    \centering
    \includegraphics[width=0.8\linewidth]{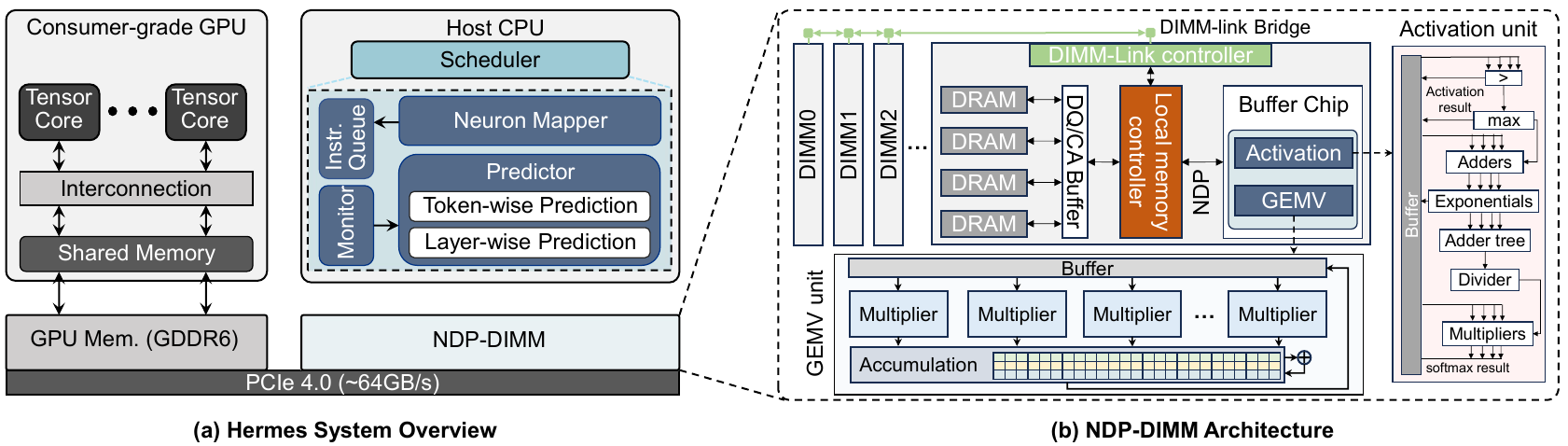}
    \vspace{-0.3cm}
    \caption{Overview of our proposed Hermes System. (a) Hermes augments GPU memory with NDP-DIMMs, and utilizes a scheduler to control the inference workflow. (b) Multiple NDP-DIMMs are connected to support LLM inference and inter-DIMM communication. }
    \label{fig:system-overview}
\vspace{-0.3cm}
\end{figure*}

\section{\name~System}\label{sec:hermes-system}
\subsection{System \& Architecture Overview}

\subsubsection{Architecture}

\fig\ref{fig:system-overview}a illustrates the overview of \name. \name~augments consumer-grade GPU with NDP-DIMMs to achieve low-budget, high-performance inference system for LLMs. 


\textbf{Consumer-grade GPU:} For LLM inference, we only use one accessible and budget-friendly consumer-grade GPU. Despite limited graphic memory, it has ample computing units, like tensor cores, for high-performance parallel processing. It also features high-speed GDDR memory with superior bandwidth. For instance, the NVIDIA RTX 4090 with 24GB GDDR6 provides 82.6 TFLOPS, 1321 Tensor TOPS, and 936 GB/s bandwidth, making it suitable for LLM inference. Hermes uses a single GPU to efficiently execute hot neurons.

\textbf{NDP-DIMM:}
Given that cold neurons are randomly activated, all data stored on each DIMM should be accessible to its own NDP units. \update{Meanwhile, DIMM is required to support the normal data access from GPU for hot neuron transmission.} Therefore, as illustrated in \fig \ref{fig:system-overview}b, we have chosen the center buffer-based NDP-DIMM design\cite{alian2018application,cong2017aim,ke2020recnmp,kwon2019tensordimm}, which allows the processing unit to access all data in its own DIMM. 
\update{The center buffer-based NDP DIMM design also complies with the normal memory access as the newly added units will not influence the memory access function supported by the local memory controller \cite{alian2018application, ke2020recnmp}.}
Here, we detail the microarchitecture of our NDP-DIMM design. To facilitate typical operations in LLMs and potential inter-DIMM data moving, each NDP-DIMM is equipped with GEMV units, activation units, and DIMM-links~\cite{zhou2023dimm}.

\textit{GEMV Unit}: 
The GEMV unit reads data from the DRAM cell and the center buffer, performing the GEMV computation associated with cold neurons. 
\update{To support batched inference and fully utilize the bandwidth achievable within the DIMM center buffer, each GEMV unit contains 256 multipliers.}
Each multiplier is responsible for 128-bit multiplication \update{in a typical bit-serial manner~\cite{devaux2019true}}, a reduction tree-based accumulator, and a 256 KB buffer. During computation, each multiplier computes eight FP16 values simultaneously, and the accumulator is responsible for the addition of partial sums with data dependencies. The buffer stores the intermediate result generated by LLM layers. 


\textit{Activation Unit}: The activation unit is designed to support the necessary non-linear functions, such as softmax and ReLU operation for LLM inference. 
This unit is composed of 256 FP16 exponentiation units, 256 FP16 addition units, and 256 FP16 multiplication units, in addition to a comparator tree, an adder tree, and a divider. 


\textit{DIMM-link}: 
Due to the input-specific nature of the activated neurons, it is necessary to adjust the neuron mapping in multiple NDP-DIMMs to further ensure the load balance of computation in the DIMMs. Therefore, we adopt DIMM-link~\cite{zhou2023dimm} to achieve inter-DIMM communication with a bandwidth of 25 GB/s. Each DIMM-link employs bidirectional external data links between DIMMs, facilitating efficient point-to-point data transfers. The DIMM-link controller and bridge enable high-speed neuron redistribution between DIMMs.
\update{Compared to relying on the host for inter-DIMM data movement, using DIMM links provides over a 62$\times$ speedup for data transfer with negligible hardware overhead.
For example, when running OPT-66B, the introduction of DIMM-link effectively reduces the migration overhead for cold neurons from 5.3\% of total time to below 0.2\% .}

\textbf{Scheduler}:
During LLM inference, the scheduler in the host CPU redistributes neuron computation tasks to the GPU and NDP-DIMMs. The scheduler primarily comprises two components: a lightweight predictor and a neuron mapper, \update{which are both implemented by software}. In addition, the scheduler includes a monitor that gathers runtime information to assist the predictor and an instruction queue that triggers instructions for the GPU and NDP-DIMMs. With the help of the monitor, the lightweight predictor leverages token-wise similarity and layer-wise correlation patterns to accurately predict neuron activity. Based on the prediction results, the neuron mapper assigns hot and cold neurons to DIMMs and GPU memory, respectively, and it also dynamically adjusts the neurons' placement to ensure efficient inference on both the GPU and NDP-DIMMs. The subsequent sections will provide detailed descriptions of these two components. 

\textbf{\update{Programming Interface}}:
\update{We use a standard programming model, PIM-SYCL \cite{kim2023samsung}, to compile the heterogeneous platform. Unified memory programming \cite{zhao2024pim, nvidia_unified_memory} allows data to be transferred implicitly between heterogeneous memory devices, enabling cooperative processing on GPU and NDP-DIMMs. Additionally, \name~provides a set of extra NDP commands, such as MAC and softmax, to support various operators in LLMs. Taking GEMV computation as an example, the NDP-DIMM computations can be invoked through the memory command interface by sending a series of MAC commands. On the GPU side, the corresponding computations are triggered through APIs like cudaLaunchKernel. }


\begin{figure}[t]
    \centering
    \includegraphics[width=\linewidth]{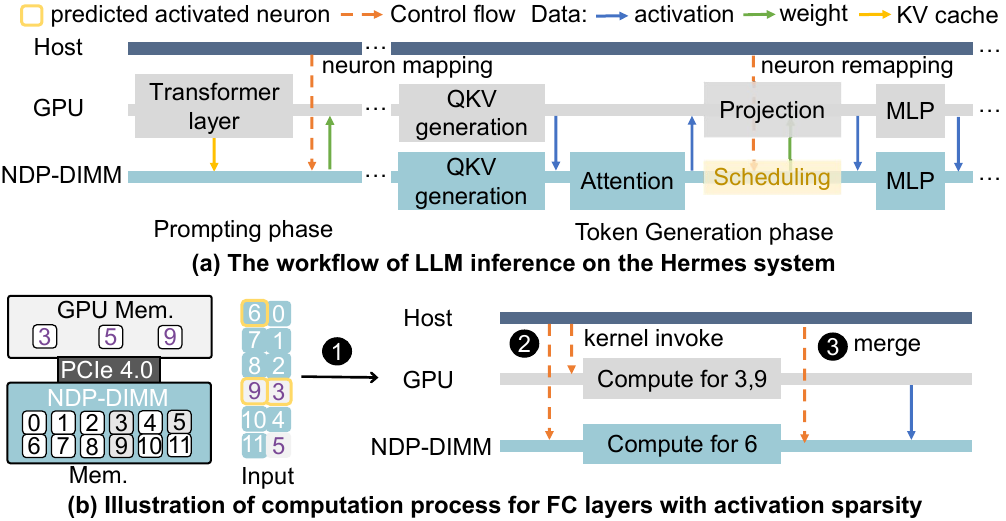}
    \vspace{-0.3cm}
    \caption{Workflow of Hermes system (a) The whole workflow of LLMs inference on the \name~system. (b) Illustration of computation process for FC layers with activation sparsity. The block with a number in the Mem. means one neuron's weight.}
    \label{fig:workflow}
\vspace{-0.3cm}
\end{figure}

\subsubsection{Workflow}

The workflow for LLM inference within the \name~system is depicted in \fig \ref{fig:workflow}a. \update{Given the significant computational demands, the entire prompting stage is processed on the GPU, adhering to a traditional offloading strategy~\cite{sheng2023flexgen}.} During this stage, the host scheduler records neuron activity for future scheduling optimization. Upon completing the prompting stage, only the selected hot neurons are loaded back into GPU memory. The offline partition of hot and cold neurons will be further detailed in Section \ref{sec:offline}.
In the token generation stage, for each transformer layer, the QKV generation is collaboratively completed by GPU and NDP-DIMMs. The output of QKV generation will be collected in the NDP-DIMMs for further attention computation. 
The memory bandwidth-intensive nature of attention computation~\cite{yu2022orca, park2024attacc} makes it ideal for execution on NDP-DIMMs, which benefit from the abundant internal bandwidth. Additionally, transferring attention computation to NDP-DIMMs helps save the limited GPU memory by eliminating the need for storing KV cache.
Since the projection layer cannot utilize the activation sparsity, it is handled solely by the computation-efficient GPU. During the projection computation, as the DIMMs are entirely idle, the host takes advantage of this period to dynamically reconfigure the hot/cold partitions and redistribute neurons across DIMMs based on the prediction results, which will be detailed in the \sec \ref{sec:partition-design} and \ref{sec:cold-neuron-mapping}. Then, similar to the QKV generation, MLP is offloaded to both GPU and NDP-DIMMs. Finally, the output of each transformer layer is reduced in the NDP-DIMMs. 



\fig \ref{fig:workflow}b illustrates the computation process for FC layers (for both QKV generation and MLP block) with activation sparsity. Specifically, it includes three steps.
After completing the related prediction, the host CPU determines the computation allocation for both the GPU and NDP-DIMMs based on the location of the activated neurons.
\update{Once the neuron mapping is determined, the host CPU invokes APIs for both the GPU and NDP-DIMMs to load data and perform computation. For example, the host CPU uses ``cudaLaunchKernel" to launch GPU kernels for GEMM and GEMV operations. To ensure correctness, the host CPU inserts barriers for the GPU and NDP-DIMMs to synchronize their computations. Once the DIMMs and GPU complete their computations, a merge kernel is invoked on the NDP-DIMMs side to gather the results from both sources.} This method is advantageous for two reasons. Firstly, as the GPU generally finishes computation tasks more quickly owing to its superior computation capability, the latency in transferring data from the GPU to DIMMs can be hidden by the DIMMs' computation, thus not penalizing the overall system runtime. Secondly, with the attention computation occurring on NDP-DIMMs, merging the QKV generation outcomes on the NDP-DIMMs side minimizes the additional data transfer overhead.

\begin{table}
\vspace{-0.3cm}
\caption{Terminology for the offline partition solver. } 
\label{tab:terminology}
\scriptsize{
\centering
\begin{tabular}{c|p{7.2cm}}
\hline
\rowcolor{black!10}
\multicolumn{2}{c}{\textbf{\textit{Parameters - predetermined or offline profiled}} }\\
\hline
$\mathbb{L}$ & All layers \\
$\mathbb{N}$ & All neurons \\
$\mathbb{D}$ & All NDP-DIMMs \\ 
$f_{i}$ & Activation frequency of neuron $i$ \\
$N_{l}$ & Neuron in layer $l$ \\
$M_{i}$ & The memory space required by neuron $i$ \\
$T_{sync}$ & The time required for one synchronization \\
$T_{l}^{j}$ & The time for computing one neuron in layer $l$ on processing $j$ \\
$S_{j}$ & The storage size for processing unit $j$ \\
\hline
\rowcolor{black!10}
\multicolumn{2}{c}{\textbf{\textit{Binary Variables - needed to be solved}} } \\
\hline
\multirow{2}{*}{$x_{il}^{j}$} & Whether neuron $i$ in layer $l$ is placed on processing unit $j$ \\
& $x_{il}^{j}=1$ means the neuron $i$ in layer $l$ is placed on processing unit $j$\\
\hline 
\end{tabular}
}
\vspace{-0.3cm}
\end{table}

\subsection{Offline Neuron Mapping} \label{sec:offline}
Since NDP-DIMMs and GPU are responsible for the computational load of the neurons stored in them, the predetermined mapping for each neuron's location greatly influences the inference efficiency. However, due to the huge neuron mapping space (e.g., more than $2^{1000}$ for LLaMA-7B), solely relying on online mapping solutions is impractical and will contribute to considerable performance degradation. Therefore, in the belief that ``hot" and ``cold" neurons are partly attributed to the pretrained LLM's nature~\cite{song2023powerinfer, song2024prosparse, zheng2024learn, song2024turbo}, we utilize the offline profiled information to deduce the initial offline neuron mapping. It alleviates the adjustment cost of subsequent online partition and scheduling during inference. Please note that the optimal initial mapping denotes the mapping that can be found during the offline stage, which will be adjusted during runtime.



To determine the optimal location for each neuron \update{that minimizes the inference latency using our heterogeneous system}, we formalize the mapping issue as an integer linear programming problem (ILP). In particular, we analyze several factors, including each neuron's activated frequency, computational overhead, memory usage, and synchronization delays, to model the inference performance of the \name~system. \update{To gather these factors accurately, we test the model on popular datasets such as C4~\cite{raffel2020exploring} and Pile~\cite{gao2020pile} with 128 samples}, and also employ an execution monitor in the host CPU to record during inference. The notation for solving the optimal offline neuron placement problem is summarized in Table \ref{tab:terminology}.

\textbf{Objective function.}
The objective of the optimal neuron mapping is to minimize the total inference latency, as shown in Equation \ref{eq1}. Since the execution of each layer involves both GPU and NDP-DIMMs, the total execution time is determined by the longer duration of the GPU and NDP-DIMMs execution times. For NDP-DIMMs, the single-layer execution time is the longest execution time among all DIMM modules, as shown in Equation \ref{eq2}. For the GPU, the single-layer execution time includes both computation time and extra synchronization overhead, while the synchronization overhead includes that of fetching input activation data from the DIMM and sending the computation results back to the DIMM to trigger the merge kernel. Hence, as illustrated in Equation \ref{eq3}, the total GPU execution time also includes twice the single-direction synchronization overhead. 

{
\setlength\abovedisplayskip{0pt}
\setlength\belowdisplayskip{0pt}
\begin{equation}
    \min \textstyle \sum_{l} max (T_{GPU-l}, T_{DIMM-l}), \quad \forall l \in \mathbb{L} \label{eq1}
\end{equation}
\begin{equation}
     T_{DIMM-l} = max (T_{dimm-jl}), \quad \forall j \in \mathbb{D} \label{eq2}
\end{equation}
\begin{equation}
    T_{GPU-l} = T_{compute-GPU-l} + 2\cdot T_{sync} \label{eq3}
\end{equation}
}

The computation times for a single layer on both the GPU and NDP-DIMMs depend on the number of activated neurons located in each device. Let $T_{l}^{GPU}$ represent the time required to compute a single neuron on the GPU. Consequently, the computation time for a single layer on the GPU is the product of the number of activated neurons in the GPU memory and the time taken to compute each neuron, as illustrated in Equation \ref{eq4}. Similarly, the single-layer computation time for each NDP-DIMM is demonstrated in Equation \ref{eq5}.

{
\setlength\abovedisplayskip{0pt}
\setlength\belowdisplayskip{5pt}
\begin{align}
    T_{compute-GPU-l} = T_{l}^{GPU} \cdot \textstyle \sum_{i} f_{i}\cdot x_{il}^{GPU}, \quad \forall i \in \mathbb{N} \label{eq4}\\ 
    T_{dimm-jl} = T_{l}^{DIMM} \cdot \textstyle \sum_{i} f_{i} \cdot x_{il}^{dimm-j}, \quad \forall i \in \mathbb{N} \label{eq5}
\end{align}
}

\textbf{Constraints.}
The offline optimal neuron placement issue must adhere to the conditions listed in \eq \ref{eq6} and \ref{eq7}, which limit the memory space occupied by neurons not to exceed the available memory size of each DIMM and GPU.

{
\setlength\abovedisplayskip{0pt}
\setlength\belowdisplayskip{5pt}
\begin{align}
   \textstyle \sum_{l} M_{i} \cdot x_{il}^{GPU} \le S_{GPU}, \quad \forall l \in \mathbb{L} \label{eq6}\\ 
   \textstyle \sum_{l} M_{i} \cdot x_{il}^{dimm-j} \le S_{dimm-j}, \quad \forall l \in \mathbb{L} \label{eq7}
\end{align}
}

Consequently, we employ the open-sourced optimization solver, PulP~\cite{pulp-solver}, to determine the optimal offline neuron mapping.
Based on our assessment, it takes about 110 seconds to solve for the optimal neuron mapping, making it appropriate for a single offline compilation process. Before LLM inference, we initially transfer relevant hot neurons to GPU memory based on the mapping outcomes and further adjust the mapping during runtime to improve efficiency.

\subsection{Online Adjustment for Hot/Cold Neuron Partition}\label{sec:partition-design}  

Although the optimal offline neuron mapping provides an effective hot/cold partition, the input-specific nature of activation sparsity makes the hot/cold neuron partition change dynamically in practice. Our evaluation indicates that about 52\% of the initialized hot neurons exhibit varied activity during inference. Therefore, it is necessary to adjust the hot/cold neuron partition online to improve inference efficiency before neuron computation, which requires an in-advance prediction of the neuron partition. In this section, we leverage the distribution patterns of activation sparsity to create a novel lightweight predictor to guide the online adjustment of the hot/cold neuron partition.
\begin{figure}[t]
    \centering
    \includegraphics[width=\linewidth]{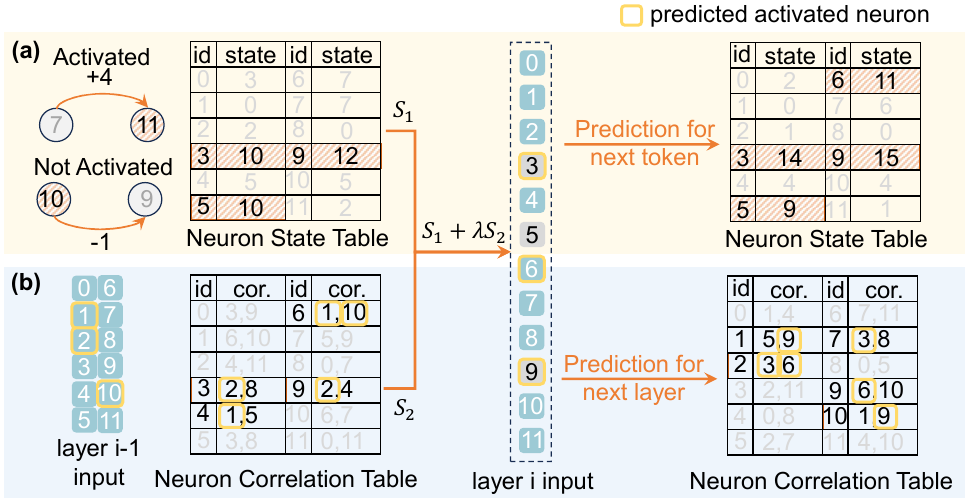}
    \vspace{-0.3cm}
    \caption{The predictor design in Hermes. (a) We are motivated to utilize the temporal locality of token generation for prediction. (b) The layer-wise correlation effectively predicts activated neurons.}
    \label{fig:predictor}
\vspace{-0.3cm}
\end{figure}

\subsubsection{Predictor Design}\label{sec:predictor-design}
Accurately forecasting activated neurons and the hot/cold neuron partition is crucial for improving inference performance. On one hand, to effectively harness activation sparsity, the \name~workflow necessitates predetermining the computation loads for both the GPU and NDP-DIMMs. On the other hand, assigning hot neurons to the GPU before computation can fully utilize the GPU’s computation capability and ease the burden on NDP-DIMMs. Nevertheless, existing MLP-based predictors~\cite{song2023powerinfer, song2024prosparse, liu2023deja} incur considerable storage and computation overhead, reducing inference efficiency. To address it, we introduce a lightweight predictor that exploits token-wise similarity and layer-wise correlation (discussed in Section \ref{sec:similar}) for accurate predictions.

\textbf{Token-wise Prediction. } The token-wise similarity suggests that the distribution of activated neurons is similar among adjacent tokens. Given that tokens are generated one by one during the token generation stage, token-wise similarity can be considered as a temporal locality of activated neurons. Inspired by well-known branch prediction strategies~\cite{smith1981study, yeh1991two, mcfarling1993combining} that also benefit from temporal locality, we propose a novel prediction strategy. As shown in \fig \ref{fig:predictor}a, we establish a neuron state table where each neuron has a 4-bit state, ranging from 0 to 15, used to predict whether the neuron will be activated. After the prefill stage, we initialize each neuron's state based on the activated frequency in the whole prefill stage. Specifically, we divide the distribution of the activated frequency into 16 stages and initialize each state accordingly. For example, if a neuron's activated frequency exceeds 90\%, its state is initialized as `15', whereas if the ratio is below 2\%, the state is set as `0'.

We update each neuron's state based on the actual activated neurons during each token generation step using a finite state machine. If a neuron is not activated, its state decreases by 1; if it is activated, its state increases by \( s \), which is set to 4 in this paper. The left part of \fig \ref{fig:predictor}a shows that, when neuron 6 is activated, the state is updated from $7$ to $11$, while the state of neuron 5 is updated from 10 to 9 as it is not activated.  

\begin{figure}
    \centering
    \includegraphics[width=0.9\linewidth]{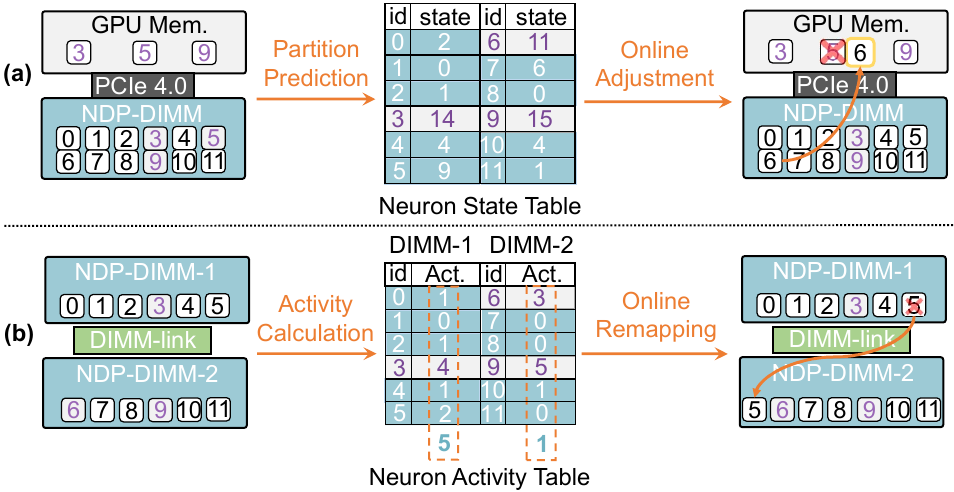}
    \vspace{-0.3cm}
    \caption{Neuron mapper design. (a) The mapper utilizes the information in the neuron state table to adjust the hot/cold neuron partition. (b) Cold neurons are remapped based on the neuron activity within a window.}
    \label{fig:mapper}
\vspace{-0.3cm}
\end{figure}

\textbf{Layer-wise Prediction. } 
Token-wise similarity alone cannot address fluctuations in neuron activity between tokens~\cite{zhang2024relu, zheng2024learn}. Therefore, we further employ layer-wise correlation to improve prediction accuracy. Insights from Section \ref{sec:similar} suggest that if neurons with high correlation in the preceding layer are activated, the activated probability for the current neuron is significantly increased. Consequently, we create a neuron correlation table to boost layer-wise prediction. As depicted in Figure \ref{fig:predictor}b, we initially offline sampled the top 2 correlated neurons from the previous layer and documented their relationships in the neuron correlation table.


Finally, we combine the token-wise and layer-wise prediction strategies to achieve accurate prediction for activated neurons during token generation. Specifically, we use $s_1$ to denote the state in the neuron state table for one neuron, and use $s_2$ to indicate the activated number of the highly correlated neurons for one neuron. To predict the activation state for such a neuron, we examine the inequation: $s_1 + \lambda\cdot s_2> T $. In this paper, we set $\lambda$ as 6, and the threshold $T$ as 15. As Figure \ref{fig:predictor} shows, following the prediction criterion, we finally activate neurons 3, 6 and 9 for subsequent computation. During context switches, token similarity may vanish, but layer-wise correlation is still available for effective prediction. Conversely, even if correlated neurons are not activated, observing neighboring tokens' activation states still helps achieve accurate prediction. Experimental result shows that the accuracy of our proposed predictor achieves 98\% using less than 1MB of memory. \update{For instance, LLaMA-7B occupies 32 layers, with each one having 4K neurons for the self-attention block and 10.5K for the MLP block. In our implementation, only 4-bit data is used to record the corresponding state of each neuron. Consequently, it only costs 232 KB for the neuron state table of LLaMA-7B.} We integrate the proposed predictor into the host CPU and store the table values in the last level cache for fast prediction.

\subsubsection{Online Adjustment guided by Predictor}
Given their ample memory capacity, instead of mapping only cold neurons, we store all the weight parameters on DIMMs. Thus, we only need to reload the actual hot neurons onto GPU memory to achieve online adjustment. The neuron state in our proposed predictor effectively represents the activity of each neuron. Specifically, as shown in the \fig \ref{fig:mapper}a, once the neuron state exceeds a certain threshold $T_h$, it can be viewed as the hot neuron. In this paper, we set the threshold $T_h = 10$. Accordingly, neurons 3, 6, and 9 are identified as hot neurons. We then use the neuron mapper to locate the corresponding hot neuron. As the hot neuron 6 is originally located on the DIMMs, an instruction is issued to copy the corresponding hot neuron to the GPU memory during the projection computation. Meanwhile, the neuron with the lowest state value (neuron 5) stored in GPU memory will be swapped out. Note that, since all neurons are stored in DIMMs, we only need to overwrite the location of the neuron to be swapped out in the GPU memory to achieve neuron swapping. In general, online neuron adjustment between GPU and NDP-DIMMs significantly improves the inference efficiency without inducing additional data transfer overhead.

\subsection{Online Remapping for Cold Neurons}\label{sec:cold-neuron-mapping}
Due to our implementation of a center buffer-based NDP-DIMM architecture, the total computation delay correlates with the count of activated neurons in each DIMM module. As shown in Equation \ref{eq2}, the total execution duration is constrained by the slowest-performing NDP-DIMM module. Hence, determining the optimal cold neuron assignment to ensure a balanced load across multiple NDP-DIMMs is crucial.
Despite using DIMM-link for inter-DIMM communication, the limited bandwidth (25GB/s) cannot afford over-frequent data exchanges between DIMMs. Therefore, we need to achieve a load balance across multiple NDP-DIMMs while minimizing the remapping of cold neurons.

\begin{algorithm}[t]
\scriptsize
    \caption{Window-based online scheduling}\label{alg:balance}

\SetKwInOut{Input}{Input}\SetKwInOut{Output}{Output}

\Input{neuron mapping $C_{j,i}$; Activity for neuron $i$ within a window $A_{i}$; Number of NDP-DIMM modules $J$;}

\emph{{\textcolor{magenta}{// Compute the number of activated neurons for NDP-DIMM $i$.}}}
$Z_{j} = \sum_{i} C_{j,i} \cdot A_{i}$ \\ 
Sort $Z$ with the descending order \\

\For{{\textcolor{blue}{int}} id = 0; $id$ $<$ J/2; id++}{
  \While{$Z_{id} \le Z_{J-id}$}{
    Find the most activated neurons $h$ in NDP-DIMM $id$\\ 
    \emph{{\textcolor{magenta}{// Remapping the most activated neurons from $id$ to $J-id$}}}
    $C_{id, h} = 0$; $C_{J-id, h} = 1$
  }
}
\end{algorithm}

The similarity between tokens inspires us to develop a novel window-based online scheduling method for remapping cold neurons. In particular, we group every five consecutive tokens into a window. Based on our observations, due to the token-wise similarity, once the optimal mapping for cold neurons is identified, the runtime variance among different NDP-DIMMs within a window is under 5\%, indicating a balanced assignment. Nevertheless, when surpassing the window size, the performance disparity among different NDP-DIMMs varies from 1.2$\times$ to 2.5$\times$. Consequently, we can leverage the neuron activity within a window to guide the remapping of cold neurons. As shown in Algorithm \ref{alg:balance}, we initially gather the activated times for each neuron $i$ within a window and calculate the total activated neurons in NDP-DIMM $j$ based on the current neuron mapping $C_{j,i}$. $C_{j,i}$ is a binary matrix that denotes if neuron $i$ is mapped on NDP-DIMM $j$. We then sort the total activated neurons for NDP-DIMMs within the window and adjust neuron mappings between DIMM pairs accordingly. Specifically, the NDP-DIMM with the largest number of activated neurons is paired with the one that has the fewest activated neurons. Finally, the most activated neurons in the NDP-DIMM pair are remapped to achieve balance. As depicted in \fig \ref{fig:mapper}b, we record the activated neurons within a window into the neuron activity table, and calculate the activity for each NDP-DIMM based on the mapping results. As the count of activated neurons in DIMM-1 exceeds that of DIMM-2, neuron 5 from DIMM-1 is remapped to DIMM-2 for load balance between the two NDP-DIMMs. This strategy offers two advantages: first, the fixed inter-DIMM communication traffic is directed to different bridges to prevent congestion; second, the greedy remapping approach can quickly achieve balance with minimal data transfer.

\section{Evaluation}

\begin{table}
    \centering
    \caption{Configuration details of NDP-DIMM.}
    \vspace{-0.3cm}
    \resizebox{\linewidth}{!}{
    \begin{tabular}{c|c|c}
    \hline
    \multicolumn{3}{c}{\textbf{NDP core}} \\
    \hline
    \multicolumn{3}{c}{Configuration: 256 multipliers, reduction tree-based accumulator, Buffer size: 256KB}\\ 
    \hline
    One NDP core per DIMM & Frequency: @ 1 GHz  & area overhead: $1.23mm^2$ per core\\
    \hline
    \multicolumn{3}{c}{\textbf{DIMM Parameters}} \\
    \hline
     \multicolumn{3}{c}{DDR4-3200, 32GB/DIMM$\times$8, \update{2 DIMMs/channel}}\\
     \multicolumn{3}{c}{4 rank/DIMM, 2 bank groups/rank, 4 bank/BG}\\
     \hline
    \multicolumn{3}{c}{\textbf{DIMM Timing}} \\
    \hline
     \multicolumn{3}{c}{tRC=76, tRCD=24, tCL=24, tRP=24, tBL=4}\\
     \multicolumn{3}{c}{tCCD S=4, tCCD L=8,tRRD S=4, tRRD L=6, tFAW=26}\\
    \hline
    \multicolumn{3}{c}{\textbf{DIMM-Link Parameters}} \\
     \hline
    \multicolumn{3}{c}{25Gb/s/Lane, 1.17 pJ/b, 8 $\times$ Lanes (25GB/s per Link)} \\
    \hline
    \end{tabular}
    }
    \label{tab:dimmcfg}
\vspace{-0.3cm}
\end{table}

\subsection{Experimental Setup}\label{sec:experimental-setup}

\subsubsection{\name~System}
The proposed \name~system integrates a single NVIDIA RTX 4090 GPU with 24GB of graphic memory \update{and 330 tensor TOPS (FP16)} to process hot neurons. Additionally, we provide 8 NDP-DIMMs, each including 32GB DDR4 memory as the extension of GPU memory. We use PCIe 4.0 to support data interaction between NDP-DIMMs and GPU memory with a bandwidth of 64GB/s. The kernel performance of the NVIDIA RTX 4090 is measured using NVIDIA Nsight Compute~\cite{nsight}. Furthermore, we develop an in-house simulator by modifying Ramulator 2.0~\cite{luo2023ramulator, ramulator2.0} to evaluate the performance efficiency of NDP-DIMM devices. For the NDP core, we implemented it in RTL and synthesized it using the Synopsys Design Compiler~\cite{synopsys.org} with the TSMC 7nm technology. \tab \ref{tab:dimmcfg} shows the configuration details of adopted NDP-DIMMs.

\subsubsection{Baseline Systems}
We selected several offloading-based inference systems, such as Huggingface Accelerate~\cite{jain2022hugging, huggingface-accelerate}, FlexGen~\cite{sheng2023flexgen}, and Deja Vu~\cite{liu2023deja}, as the baselines. FlexGen and Deja Vu are restricted to OPT models. Moreover, Deja Vu, initially optimized for LLM activation sparsity within high-performance distributed systems, has been adapted to support offloading-based serving systems. In contrast to \name, these methods depend solely on the basic host memory to expand capacity without offering additional computational resources. \update{We also provided a system (Hermes-host) that offloads cold neurons to the host CPU while handling hot neurons on GPU, demonstrating the necessity of NDP-DIMMs. Hermes-host follows the configuration in~\cite{song2023powerinfer}, which equips an Intel i9-13900K processor as the host CPU (providing a maximum bandwidth of 89.6 GB/s), and also uses a single NVIDIA RTX 4090 as the GPU for hot neurons.} Additionally, to highlight the significance of activation sparsity in boosting \name~system efficiency, we also compare \name~against a straightforward NDP-DIMM extended system (referred to as Hermes-base) that does not leverage activation sparsity in LLMs.

\begin{figure}
    \centering
    \includegraphics[width=.98\linewidth]{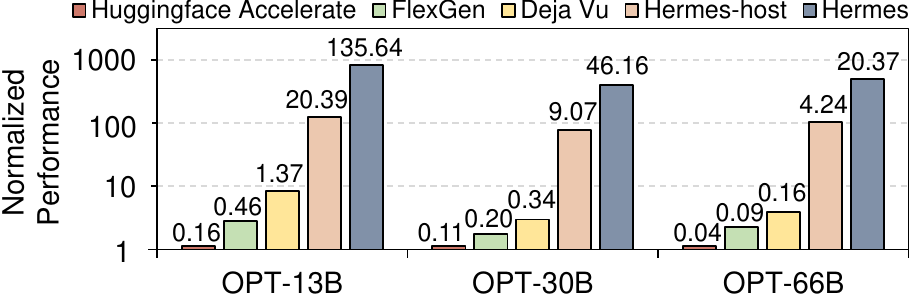}
    \vspace{-0.3cm}
    \caption{\update{Performance comparison with existing offloading-based systems.}}
    \label{fig:offloading-performance}
\vspace{-0.3cm}
\end{figure}

\begin{figure}[t]
    \centering
    \includegraphics[width=0.98\linewidth]{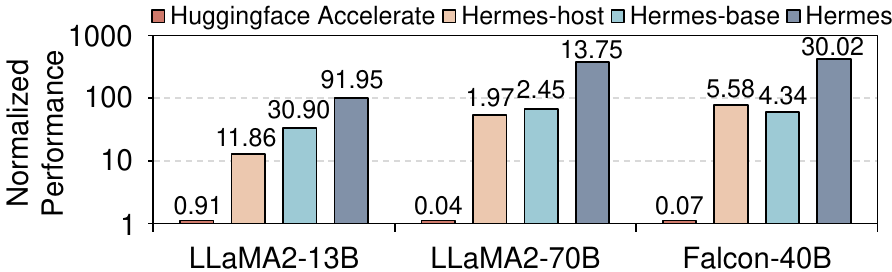}
    \vspace{-0.3cm}
    \caption{\update{The effectiveness of activation sparsity and NDP design on \name.}}
    \label{fig:base-hermes-performance}
\vspace{-0.3cm}
\end{figure}

\subsubsection{Workloads}
We chose OPT-13B, OPT-30B, OPT-66B~\cite{zhang2022opt}, LLaMA2-13B, LLaMA2-70B~\cite{touvron2023llama2}, and Falcon-40B~\cite{almazrouei2023falcon} as target models. For the OPT series models, we utilized their native ReLU activations to achieve activation sparsity. For the LLaMA2 and Falcon models, we use the open-source models\footnote{The modified LLMs can be found at \href{https://huggingface.co/SparseLLM}{https://huggingface.co/SparseLLM}, including both LLaMA2 and Falcon models} that substituted their original activation functions with ReLU~\cite{mirzadeh2023relu, zhang2024relu}. Furthermore, we added additional ReLU functions before generating QKV to achieve activation sparsity in self-attention blocks. Evaluation results show that these alterations result in negligible accuracy loss (under 1\%). \update{Furthermore, we adopt ChatGPT prompts~\cite{gpt-prompts} and Alpaca~\cite{alpaca} as the datasets to evaluate the end-to-end performance, following configurations in \cite{xue2024powerinfer, song2023powerinfer}.}

\begin{figure*}
    \centering
    \includegraphics[width=\linewidth]{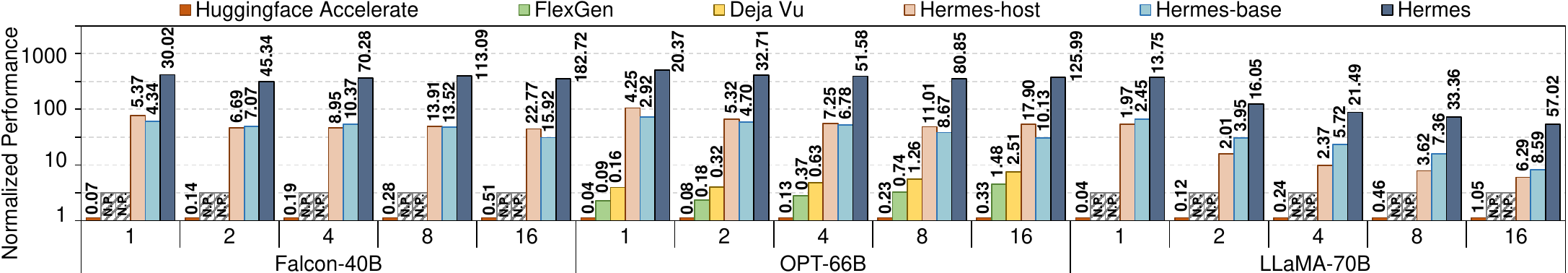}
    \vspace{-0.3cm}
    \caption{\update{End-to-end performance on different batch sizes (ranging from 1 to 16). N.P. denotes the model is not supported by the current inference system.}}
    \label{fig:batching-inference}
\vspace{-0.3cm}
\end{figure*}

\subsubsection{Evaluation Metric}
Given our focus on local deployment scenarios, we primarily optimized LLM inference with small batch sizes. We concentrated on the average number of tokens generated per second (tokens/s) to evaluate model inference efficiency. Hereafter, the number above each bar in each figure indicates the end-to-end generation speed (tokens/s). In our experiments, we used batch sizes between 1 and 16, and kept the lengths of both input and output sequences fixed at 128.



\subsection{\name~Performance}\label{sec:end-to-end}

\subsubsection{End-to-End Performance}
We begin by evaluating the end-to-end inference performance of \name~and baseline systems at a batch size of 1, which is commonly used for local deployments~\cite{cai2023medusa}. Noting that FlexGen and Deja Vu are limited to support OPT family models, we first compare \name~against existing offload-based inference systems on OPT models. \update{Additionally, we evaluate the Hermes-host and Hermes-base systems' performance across various LLMs to illustrate the necessity of NDP-DIMMs design and activation sparsity in Hermes, respectively.}

\textbf{Comparison with Offloading-based Systems. } \fig \ref{fig:offloading-performance} presents the end-to-end performances on OPT family models. Compared with the Accelerate and FlexGen systems, \name~can achieve an average $578.42 \times$ and $247.25 \times$ speedup, respectively. \name~is capable of achieving a rate of $20.37$ tokens/s for OPT-66B, which substantially surpasses current inference systems. In contrast, Deja Vu only attains an average speedup of $2.12 \times$ over FlexGen due to the necessity of loading cold neurons. The frequent data transfer on PCIe compromises the performance improvement of activation sparsity, while the expensive MLP-based predictor used in Deja Vu further diminishes its benefits. Compared to OPT-13B, \name~achieves greater performance gains on OPT-66B. This is because 80\% of the parameters in OPT-13B can be stored in GPU memory, whereas only 15\% of parameters in OPT-66B can be stored in GPU memory. This further exacerbates the data transfer overhead between host memory and GPU memory. 

\textbf{Necessity of Activation Sparsity. } We further compare \name~with the Hermes-base system, which only adopts a na\"ive NDP-DIMM extended system without utilizing activation sparsity, as shown in \fig \ref{fig:base-hermes-performance}. 
The Hermes-based system processes the FC layers on the GPU when their parameters are available, switches to NDP-DIMMs when their parameters are stored in those modules, and offloads all attention computations to NDP-DIMMs.
This approach leverages the high internal bandwidth of NDP-DIMMs and
reduces data transfer between DIMMs and GPU memory. In comparison to Huggingface Accelerate, the Hermes-base system can achieve $53.89 \times$ speedup on average, as it greatly reduces the data transfer on PCIe. By effectively leveraging activation sparsity in LLMs, Hermes outperforms the Hermes-base system with average speedups of $5.17 \times$, specifically for large models such as Falcon-40B and
LLaMA2-70B. This is due to when running large models, most layers are offloaded on the computation-limited NDP-DIMMs for the Hermes-base system. 

\update{\textbf{NDP-DIMMs instead of host CPU. } 
Experimental results in Figure \ref{fig:offloading-performance}, \ref{fig:base-hermes-performance} demonstrate the necessity of NDP-DIMMs. Hermes achieves $4.79\times$ - $7.75\times$ speedup when compared to Hermes-host. Specifically, the Hermes-host system also utilizes the hot/cold neuron partition, but computes the cold neurons on the host CPU. This approach effectively alleviates the burdensome data loading on PCIe for existing offloading-based systems. In comparison to Huggingface Accelerate and FlexGen, the Hermes-host system can achieve $62.00 \times$ and $44.96\times$ speedup on average, respectively. However, the memory bandwidth on the CPU side is significantly lower than that of NDP-DIMMs, making the Hermes-host system still far less efficient than our proposed Hermes system. }

\subsubsection{Batching Inference}

\begin{figure*}
    \centering
    \includegraphics[width=\linewidth]{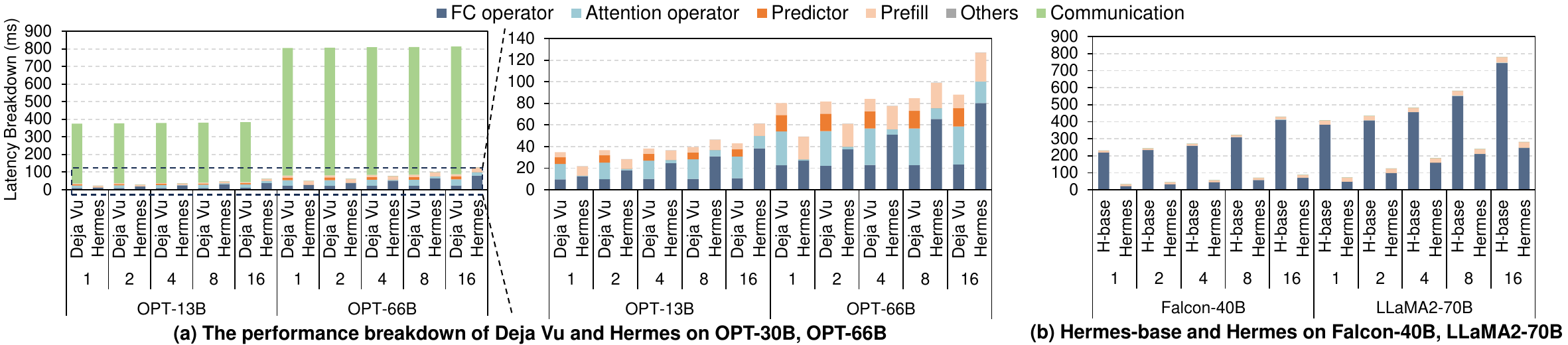}
    \vspace{-0.3cm}
    \caption{Evaluating the performance breakdown on Deja Vu, \name, and \name-base (H-base) on various LLMs with different batch sizes. }
    \label{fig:performance-breakdown}
\vspace{-0.3cm}
\end{figure*}

We also evaluate the end-to-end performance of \name~with different batch sizes. As shown in the \fig \ref{fig:batching-inference}, \name~demonstrates consistent performance improvement with the batch sizes varying from 1 to 16. Hermes attains average speedups of $148.98\times$ and $75.24\times$ for various batch sizes when compared to FlexGen and Deja Vu, respectively, offering promising support for larger batch sizes. \update{Furthermore, \name~achieves an average $7.17 \times$ speedup over \name-host for various batch sizes. As the batch size increases, the performance gap between Hermes-host and Hermes becomes more pronounced. This occurs as the consumer-grade GPU with sufficient computation capability is minimally impacted by larger batch sizes, whereas the dynamic loading overhead of cold neurons is closely tied to bandwidth. Consequently, as batch sizes grow, the limited memory bandwidth on the CPU side increasingly affects overall system performance.} The performance gap between \name~and the Hermes-base system is the smallest when the batch size is 2. This is because for \name-base, the computation capability of the NDP core can still effectively handle the corresponding computational load, and larger batches can effectively amortize the DRAM cell access overhead as weight parameters are reused by the two batches. At other batch sizes, \name~demonstrates a significant performance advantage over Hermes-base. First, at a batch size of 1, Hermes can utilize activation sparsity to significantly reduce the number of neurons that need to be activated, thereby lowering data access overhead. Second, as the batch size increases, Hermes is not constrained by the computation capability of NDP-DIMMs due to the presence of activation sparsity. 

\subsection{Ablation Studies}\label{sec:ablation-study}

\begin{figure}
    \centering
    \includegraphics[width=\linewidth]{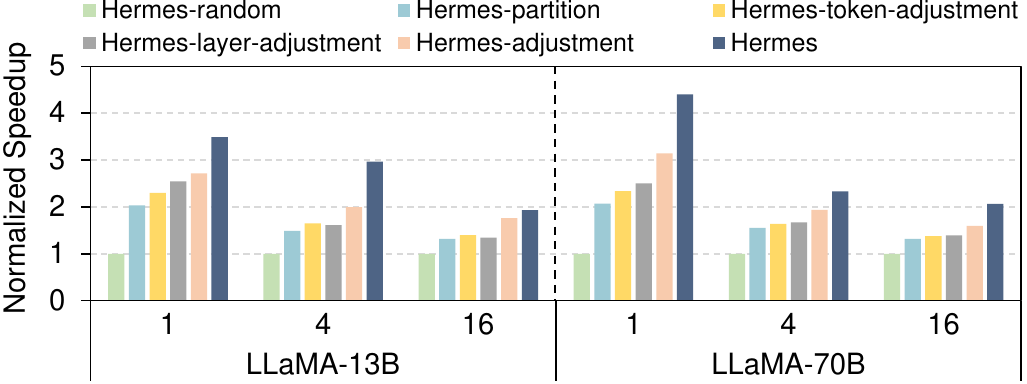}
    \vspace{-0.3cm}
    \caption{\update{Ablation study on proposed offline and online scheduling strategies.}}
    \label{fig:ablation-study}
\vspace{-0.3cm}
\end{figure}

To evaluate the scheduling strategies proposed in Section \ref{sec:hermes-system}, we compare the normalized inference latency on MLP block for different LLMs with various scheduling settings. Specifically, \name-random denotes utilizing a random offline mapper to achieve neuron placement, \name-partition denotes that it only considers the optimal offline neuron placement, \name-adjustment denotes the system that further uses online adjustment for hot/cold neuron partition, and \name~is the one that integrates all the scheduling strategies proposed in Section \ref{sec:hermes-system}. \update{Furthermore, we also explore when only adopting token-wise prediction or layer-wise prediction to guide the online adjustment of hot/cold partition, denoted as Hermes-token-adjustment and Hermes-layer-adjustment, respectively.} 

\textbf{Load Balancing with Multi-level Optimization. } \fig \ref{fig:ablation-study} shows the contributions of each component in \name~. Utilizing the offline mapper can effectively identify the frequent hot neurons, reducing the computation cost of NDP-DIMMs. As a result, \name-partition can achieve $1.63 \times$ speedup than \name-random. However, the input-specific nature of activation sparsity challenges the offline partition approach. Therefore, further adopting online adjustment for hot/cold partition (\name-adjustment) achieves $1.33 \times$ performance gains over \name-partition. Despite this, the overall execution efficiency is still constrained by the NDP-DIMMs, which possess limited computation capability. Thus, the performance of the resource-constrained NDP-DIMMs can be improved by tackling the load imbalance issues in several NDP-DIMMs. The introduced online remapping method successfully addresses this problem. As a consequence,
the fully optimized Hermes system demonstrates a $1.29 \times$ boost in performance when compared with \name-adjustment.

\update{\textbf{Benefits of Token-wise and Layer-wise Prediction.}
Compared to \name-partition which only considers the optimal offline neuron placement, \name-token-adjustment and \name-layer-adjustment can achieve $1.08\times$ and $1.11\times$ speedup, respectively, demonstrating the benefits of online adjustment. However, token-wise prediction cannot address fluctuations in neuron activity, making it inaccurate for frequent changes in hot/cold neurons. Simultaneously, layer-wise prediction only relies on the static sampled neuron correlation table to guide the online adjustment, inefficient for constant changes of online adjustment. As a result, using token-wise or layer-wise prediction only cannot effectively unleash the benefits of prediction-based online adjustment.  
}

\subsection{Performance Breakdown}\label{sec:breakdown}


\fig \ref{fig:performance-breakdown} illustrates the performance breakdown of Deja Vu, \name-base, and \name~on various LLMs. It provides detailed insights into the efficiency sources of \name.

Figure \ref{fig:performance-breakdown}a shows that while Deja Vu benefits from activation sparsity, it still requires loading cold neurons when activated, resulting in communication costs—especially PCIe data transfer—comprising about 89\% of the execution time. On the right side of Figure \ref{fig:performance-breakdown}a, we disregard the effect of communication on performance. The MLP-based predictor in Deja Vu consumes roughly 18.1\% of computation time, further reducing the gains from activation sparsity. Our lightweight predictor, in contrast, contributes less than 0.1\% to runtime overhead. Even with communication costs lowered through reusable neurons at large batch sizes, Deja Vu's performance remains inferior to Hermes.

Figure \ref{fig:performance-breakdown}b compares \name-base and \name. Without activation sparsity, \name-base incurs higher computation costs, especially as batch sizes increase, due to intensive computation on NDP-DIMMs. For example, running LLaMA2-70B offloads over 80\% of computation to NDP-DIMMs, leading to a substantial portion of the execution time being occupied by FC computation. In Hermes, token generation takes 66.40\% of execution time at batch size 1. After optimizing token generation, the prompting stage becomes the bottleneck, accounting for about 33.01\% of the overhead, limiting further inference efficiency improvements.

\begin{figure}
    \centering
    \includegraphics[width=\linewidth]{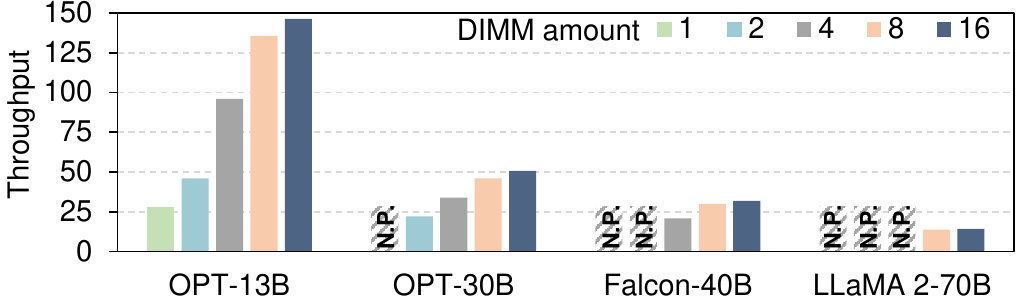}
    \vspace{-0.3cm}
    \caption{\update{Throughput of four typical LLMs with different numbers of NDP-DIMMs. N.P. denotes the model is not supported by current system.}}
    \label{fig:dimm}
\vspace{-0.3cm}
\end{figure}

\update{\subsection{Sensitivity Studies}}

\subsubsection{\update{Sensitivity analysis of the number of DIMMs}}

\update{\fig~\ref{fig:dimm} illustrates the improvement in LLM throughput as the number of NDP-DIMMs increases. We evaluated four distinct LLM models using a single batch to understand the impact of varying numbers of NDP-DIMMs, while mitigating the effect of limited computation capability. An increase in NDP-DIMMs enhances both memory size and internal bandwidth. Larger memory capacity facilitates the deployment of more extensive models; for instance, deploying Falcon-40B on Hermes necessitates a minimum of four NDP-DIMMs. Additionally, higher internal bandwidth significantly enhances end-to-end performance, addressing the bandwidth limitations that bottleneck current offloading-based systems. However, once sufficient bandwidth is achieved, further increases in the number of NDP-DIMMs do not proportionally boost throughput. For example, LLaMA2-70B exhibits similar throughput with both 8 and 16 NDP-DIMMs. Once the NDP-DIMMs surpass the GPU in performance, additional NDP-DIMMs do not yield further performance gains.}


\begin{figure}
    \centering
    \includegraphics[width=\linewidth]{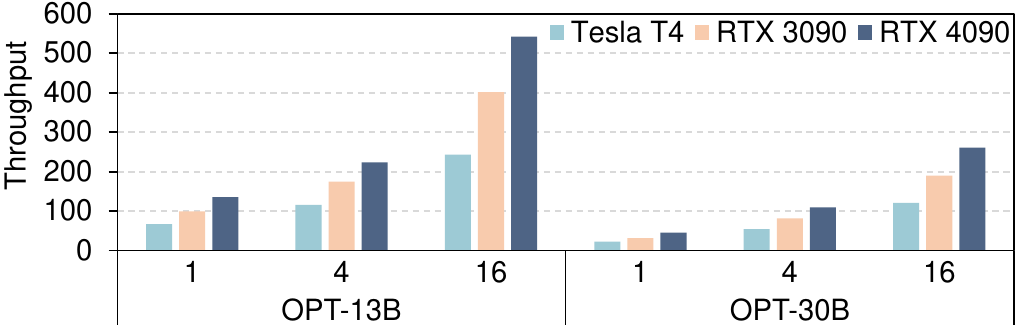}
    \vspace{-0.3cm}
    \caption{\update{Throughput of OPT-13B and OPT-30B with various GPUs, including RTX 4090, RTX 3090 and Tesla T4.}}
    \label{fig:gpu}
\vspace{-0.3cm}
\end{figure}

\subsubsection{\update{Sensitivity analysis of various GPUs}}

\update{\fig~\ref{fig:gpu} illustrates the significant impact of different GPUs on the end-to-end throughput of LLM execution. We have included two additional consumer-grade GPUs, Tesla T4 and RTX 3090, in our evaluation. Specifically, Tesla T4 offers 16GB of graphic memory, 320GB/s memory bandwidth, and 65 tensor TOPS (FP16), whereas RTX 3090 provides almost the same graphic memory and bandwidth as RTX 4090, but with 142 tensor TOPS (FP16). Overall, \name~with RTX 4090 achieves an average throughput improvement of $2.02\times$ and $1.34\times$ compared to \name~with Tesla T4 and RTX 3090, respectively. The data loading cost for RTX 3090 is nearly identical to that of RTX 4090. However, RTX 3090 spends more time on prefill and hot neuron computations due to its weaker computation capability. Tesla T4, with its smaller graphic memory and lower memory bandwidth compared to RTX 3090, is inefficient for data loading. Consequently, the choice of GPU device is crucial for optimizing \name~performance.}

\subsubsection{\update{Design Space Exploration for NDP-DIMMs}}
\begin{figure}
    \centering
    \includegraphics[width=\linewidth]{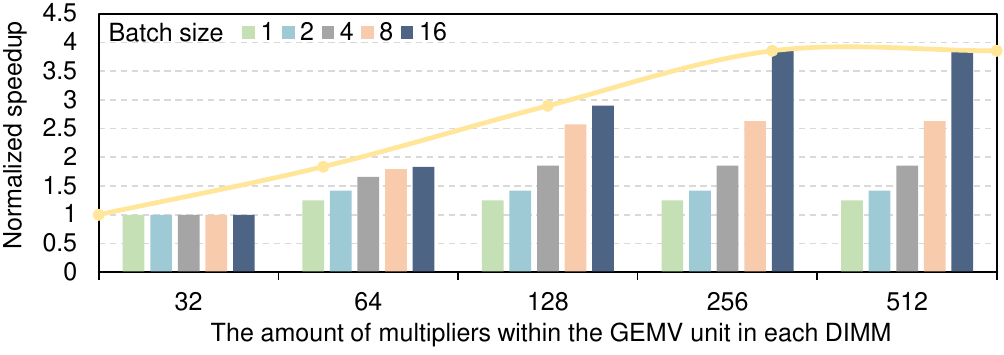}
    \vspace{-0.3cm}
    \caption{\update{Design Space Exploration for NDP-DIMMs with different number of multipliers in each GEMV unit.}}
    \label{fig:gemv}
\vspace{-0.3cm}
\end{figure}

\update{
\fig~\ref{fig:gemv} highlights the impact of increasing the number of multipliers within a GEMV unit per DIMM on LLM inference performance, especially with larger batch sizes. We varied the number of multipliers within a GEMV unit from 32 to 512, thereby enhancing computation capability by 16$\times$. For OPT-13B with a batch size of 1, performance stabilizes once 64 multipliers are reached, as further computation capability yields minimal gains. In contrast, with a batch size of 16, performance continuously improves with additional multipliers, achieving up to a $3.86\times$ speedup. This difference arises because memory bandwidth limits performance for smaller batch sizes due to lower arithmetic intensity, while computation capability becomes the bottleneck with larger batch sizes. To optimize the balance between hardware overhead and performance across various batch sizes, we selected 256 multipliers within the GEMV unit per DIMM.
}

\subsection{Comparison with High-Performance System}\label{sec:comparison-high-performance}

\begin{figure}[t]
    \centering
    \includegraphics[width=\linewidth]{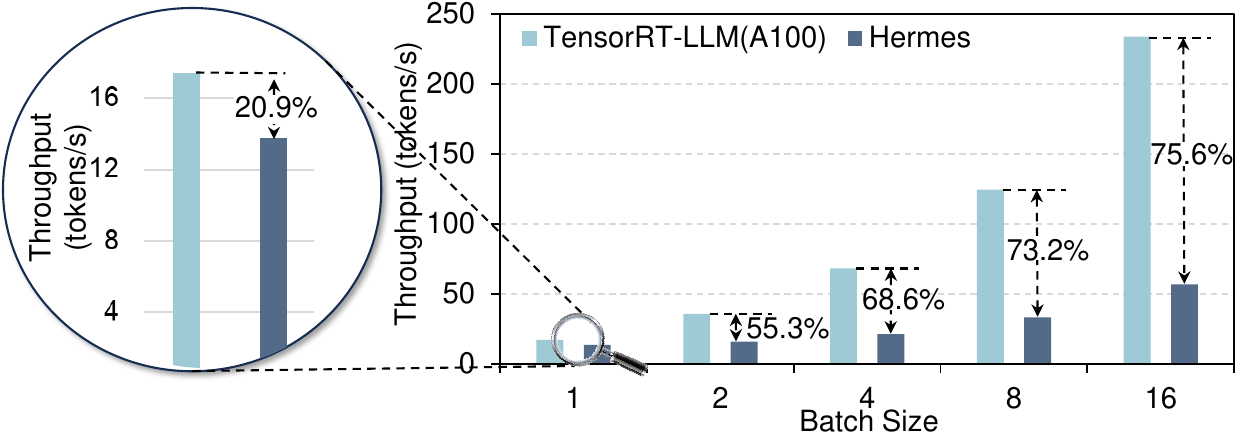}
    \vspace{-0.3cm}
    \caption{Comparison with TensorRT-LLM on LLaMA2-70B.}
    \label{fig:trt-llm-comparison}
\vspace{-0.3cm}
\end{figure}

This section discusses the performance gap between our budget-friendly LLM inference system Hermes and state-of-the-art high-performance serving system TensorRT-LLM~\cite{tensorrt-llm}. We kept the input and output sequence lengths set at 128. To handle LLaMA2-70B with a batch size of 16, TensorRT-LLM requires five NVIDIA A100-40GB-SXM4 GPUs. In contrast, 
\name~operates with only one NVIDIA RTX-4090 GPU and affordable NDP-DIMMs. Figure \ref{fig:trt-llm-comparison} displays the performance comparison between TensorRT-LLM and Hermes. For a batch size of 1, Hermes achieves 79.1\% inference efficiency of TensorRT-LLM. Even at a batch size of 16, Hermes retains 24.4\% inference efficiency of TensorRT-LLM. Despite this, Hermes is far more economical than TensorRT-LLM, which is equipped with 5 NVIDIA A100-40GB-SMX4 GPUs. Specifically, Hermes only costs approximately \$2,500, whereas TensorRT-LLM requires \$50000 to support LLaMA2-70B. Hermes provides efficient and low-budget LLM inference for local deployments. 

\section{Related Works}\label{sec:related-works}

\subsection{LLM Inference with PIM}\label{sec:related-works-PIM}

Given that LLM inference is primarily memory bandwidth-bound, accelerating it with processing in memory (PIM) is a natural choice~\cite{li2020hitm,zhai2023star,zhu2023processing}. AttAcc!~\cite{park2024attacc} utilizes a hybrid architecture of HBM-PIM and xPU (GPU/TPU), offloading the attention computation to HBM-PIM. NeuPIMs~\cite{heo2024neupims} and IANUS~\cite{seo2024ianus} address the compatibility issue between PIM functionality and regular memory access by adopting dual buffers and incorporating additional control units, respectively. They optimize the design of HBM-PIM to support both processing and memory access simultaneously, utilizing PIM and xPU collaboration for LLM inference acceleration. SpecPIM~\cite{li2024specpim}, on the other hand, targets speculative LLM models with a multi-device architecture, where each device includes an xPU and multiple HBM-PIM chips. However, these works are all designed for server-grade devices (such as H100) and rely on expensive HBM-PIM for LLM inference acceleration, making them unsuitable for local deployment with a limited budget. 

\subsection{LLM Acceleration with Activation Sparsity}\label{sec:related-works-sparsity}


The promising activation sparsity in deep learning models motivates researchers~\cite{zheng2023pit, cui2023optimizing,liu2024drift} to further improve their inference efficiency, especially for LLMs. Deja Vu~\cite{liu2023deja} utilizes the activation sparsity to reduce the memory access on the unified memory of multiple server-grade GPUs. However, it still requires storing all parameter data in GPU memory, failing to reduce GPU storage overhead. Powerinfer~\cite{song2023powerinfer} introduces a CPU-GPU hybrid system to achieve activation sparsity-based LLM inference. It stores hot neurons in GPU memory and uses GPU tensor cores for the corresponding computations while offloading cold neurons in CPU memory and utilizing the CPU as a computing unit. However, the CPU-side memory bandwidth is significantly lower than that in the GPU, making CPU-side computation a bottleneck. Overall, existing systems do not fully exploit the advantages of activation sparsity.
\section{Conclusion}\label{sec:conclusion}


In this paper, we propose an innovative and affordable inference system, Hermes, that utilizes NDP-DIMMs to enhance both the memory capacity and processing capability of consumer-grade GPUs. We partition the billion-scale weight parameters within LLMs into hot/cold neurons. Specifically, we map hot neurons to computation-efficient but storage-limited consumer-grade GPUs, while offloading cold neurons to storage-ample but computation-limited NDP-DIMMs, to fully leverage their advantages. To further improve the inference efficiency on \name, we propose a lightweight predictor to assist the online partition for hot/cold neurons and adopt window-based online scheduling to achieve load balance across multiple NDP-DIMMs. Compared with existing high-performance inference systems, \name~can achieve competitive inference efficiency with approximately 5\% budget. 

\section{Acknowledgments}
We sincerely thank the anonymous reviewers for their insightful suggestions. This work was partially supported by the National Key R\&D Program of China (Grant No. 2023YFB4404400) and the National Natural Science Foundation of China (Grant No. 62222411, 62204164). Ying Wang is the corresponding author (wangying2009@ict.ac.cn).

\bibliographystyle{plain}
\bibliography{references}

\end{document}